\documentclass[preprint]{revtex4-1}
\newcommand{\ST}{{\it Sim2L}}
\newcommand{\STs}{{\it Sim2Ls}}
\pdfoutput=1
\usepackage[lofdepth,lotdepth]{subfig}
\usepackage{graphicx}
\usepackage{natbib}
\usepackage{float}
\usepackage{amsmath}
\usepackage{amssymb}
\usepackage{epsfig}
\usepackage{subeqnarray}
\usepackage{color}
\usepackage{bbold}
\usepackage{dsfont}
\usepackage{tikz}
\usepackage{cancel}
\usetikzlibrary{positioning,calc}
\usepackage{hyperref}
\usepackage[final]{changes}

\begin{document}
\title{Sim2Ls: FAIR simulation workflows and data}

\author{Martin Hunt}
\affiliation{Halıcıoğlu Data Science Institute, University of California at San Diego, La Jolla, California, 92093, USA}

\author{Steven Clark}
\affiliation{San Diego Supercomputer Center, University of California at San Diego, La Jolla, California, 92093, USA}

\author{Daniel Mejia}
\affiliation{Network for Computational Nanotechnology, Purdue University, West Lafayette, Indiana 47907, USA}

\author{Saaketh Desai}
\affiliation{Center for Integrated Nanotechnologies, Sandia National Laboratories, Albuqerque, New Mexico 87123, USA}

\author{Alejandro Strachan}
\affiliation{School of Materials Engineering and Network for Computational Nanotechnology, Purdue University, West Lafayette, Indiana 47907, USA}

\affiliation{Corresponding author: strachan@purdue.edu}

\date{\today}
\begin{abstract}

Just like the scientific data they generate, simulation workflows for research should be findable, accessible, 
interoperable, and reusable (FAIR). However, while significant progress has been made towards FAIR data, 
the majority of science and engineering workflows used in research remain poorly documented and often unavailable,
involving {\it ad hoc} scripts and manual steps, hindering reproducibility and stifling progress. 
We introduce \STs{} (pronounced simtools) and the \ST{} Python library that allow developers to create and share  
end-to-end computational workflows with well-defined and verified inputs and outputs. The \ST{} library makes 
\STs, their requirements, and their services discoverable, verifies inputs and outputs, and automatically
stores results in a globally-accessible simulation cache and results database. This simulation ecosystem is available 
in nanoHUB, an open platform that also provides publication services for \STs{}, a computational environment for developers 
and users, and the hardware to execute runs and store results at no cost. 
We exemplify the use of \STs{} using two applications and discuss best practices towards FAIR simulation
workflows and associated data. 

\end{abstract}

\maketitle
\section{Introduction}

Scientific progress is based on the ability of researchers to independently reproduce published results and verify
inferences \cite{baker2016reproducibility,goodman2016does}.
These results are nearly universally obtained via complex, multi-step, workflows involving experiments and/or simulations
with multiple inputs, data collection, and analysis. It is often the case that, even when the authors 
carefully document their procedures, reproducing published results requires a significant investment of time
even for experts. This is true both in experimental and computational work, it slows down progress and results
in wasted resources.
A related issue hindering innovation is the fact that the majority of the data generated during research is not made
available to the community and the fraction that is used in publications, generally skewed, is often not findable or queryable. 
This is particularly problematic with the increasingly important role machine learning is playing in physical sciences and 
engineering \cite{butler2018machine,himanen2019data}.
Guidelines to making data findable, accessible, interoperable, and reusable (FAIR) have been put forward \cite{wilkinson2016fair} and a 
variety of concrete efforts to tackle these issues have been launched in recent years. 
Examples in the physical sciences range from open and queryable repositories of materials properties, 
both computational and experimental \cite{saal2013materials,curtarolo2012aflow,blaiszik2019data,jain2016research,o2016materials},
to publications devoted to scientific data \cite{naturedata2021} as well as infrastructure to publish and share models \cite{OpenKIM,strachan2010cyber}. 

FAIR principles apply not just to scientific data but also to research workflows used to generate them, 
this is particularly true for computational workflows where documentation, automatization, and reproducibility are easier than in experiments.
\cite{lamprecht2020towards}
Growing interest in making workflows available are reflected by the increasing popularity of Git repositories 
\cite{spinellis2012git} and Jupyter \cite{kluyver2016jupyter}.
Notable examples of reproducible workflows include {\it ab initio} calculations performed in the Materials Project 
\cite{jain2011high}, openKIM property calculations \cite{karls2020openkim}, osteoarthritis image processing \cite{bonaretti2020pykneer}. 
In addition, several publishers require either data availability statements or all data and codes to be made
available \cite{scimag}; some have also developed lists of suggested repositories, see, for example, \cite{data_2019}.
Despite these laudable efforts, the majority of research workflows used in published research are described 
in incomplete terms and using technical English as opposed to using specialized tools. Furthermore they often involve ad hoc 
analysis scripts and manual steps that conspire against automation and reproducibility. 
This is in part due to the lack of general tools for the development and publication of computational tools with well defined, 
verifiable, and discoverable inputs and outputs and the automatic storage of results.

\begin{figure}[htp]
    \centering
    \includegraphics[width=10cm]{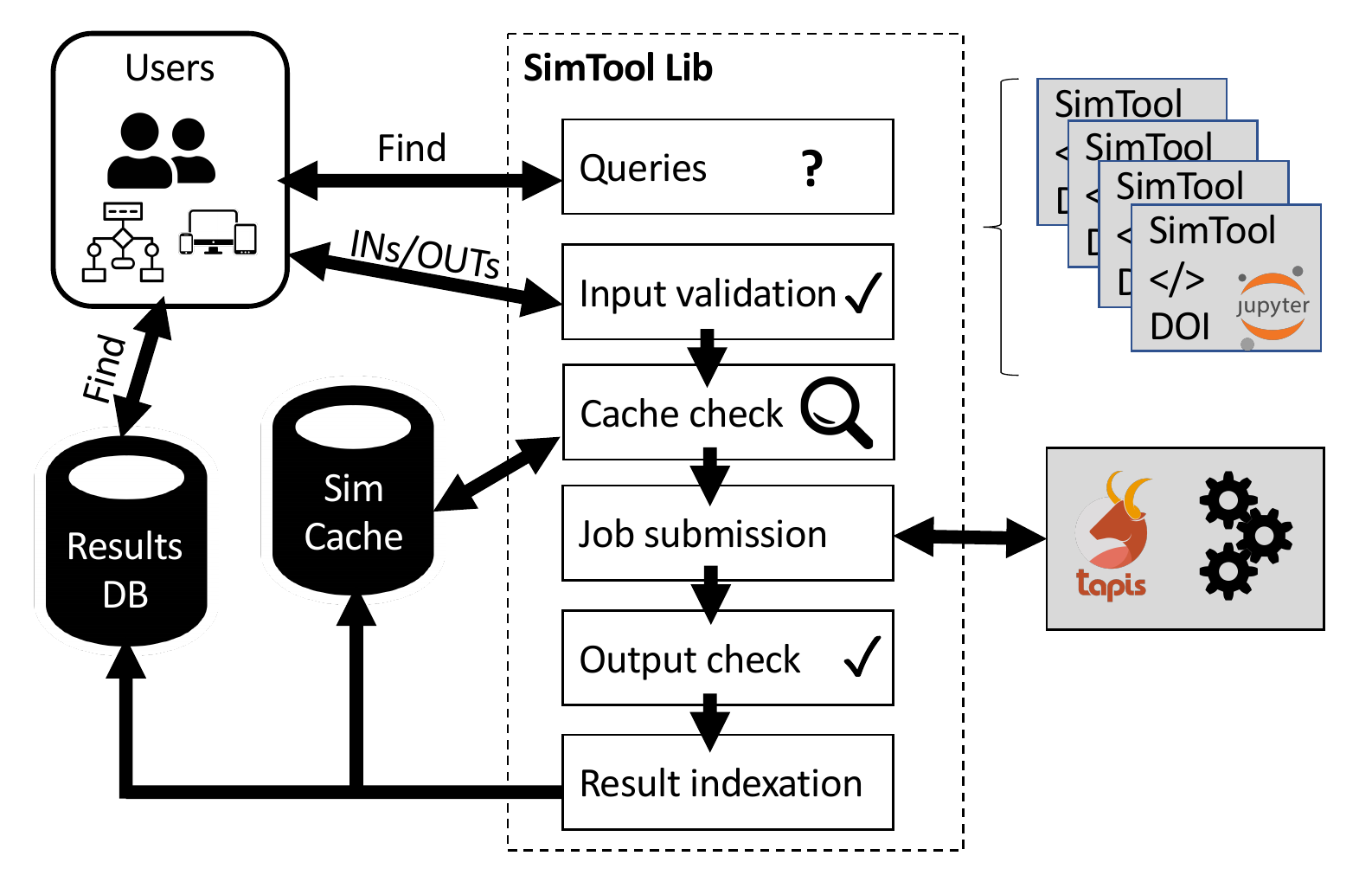}
    \caption{Simtool Ecosystem Diagram}
    \label{fig:SimtoolEcosystem}
\end{figure}

To address these gaps we introduce {\it \STs{}}, a library to create and share end-to-end computational workflows with
verified inputs and outputs, see Fig. \ref{fig:SimtoolEcosystem} for a schematic representation of the ecosystem. 
These workflows have verified inputs and outputs, could launch large-scale simulations in high-performance computing resources, 
employ a simulation cache to re-use previous runs, and index results in a database to enable querying.
The \ST{} library is available via the US National Science Foundation's nanoHUB \cite{strachan2010cyber} which also 
provides services for workflow publication, free and open online simulations of published {\it \STs{}},  
automatic caching of simulation runs, and indexing of the outputs in a queryable database. This ecosystem
makes \STs{}, their services, requirements, and the results they produce FAIR. The \ST{} library is available at
\url{https://github.com/hubzero/simtool} and its documentation at \url{https://simtool.readthedocs.io/en/stable/}.
The remainder of this paper discusses the elements of a \ST{} and the \ST{} library (Section \ref{sec:SimTool}), 
provides examples of their use (Section \ref{sect:examples}), followed by general \ST{} design guidelines for 
developers (Section \ref{sec:Discussion}) and  conclusions. 

\section{Sim2Ls and the Sim2L library}
\label{sec:SimTool}

\subsection{Elements of a Sim2L}

\begin{figure}[htp]
    \centering
    \includegraphics[width=15cm]{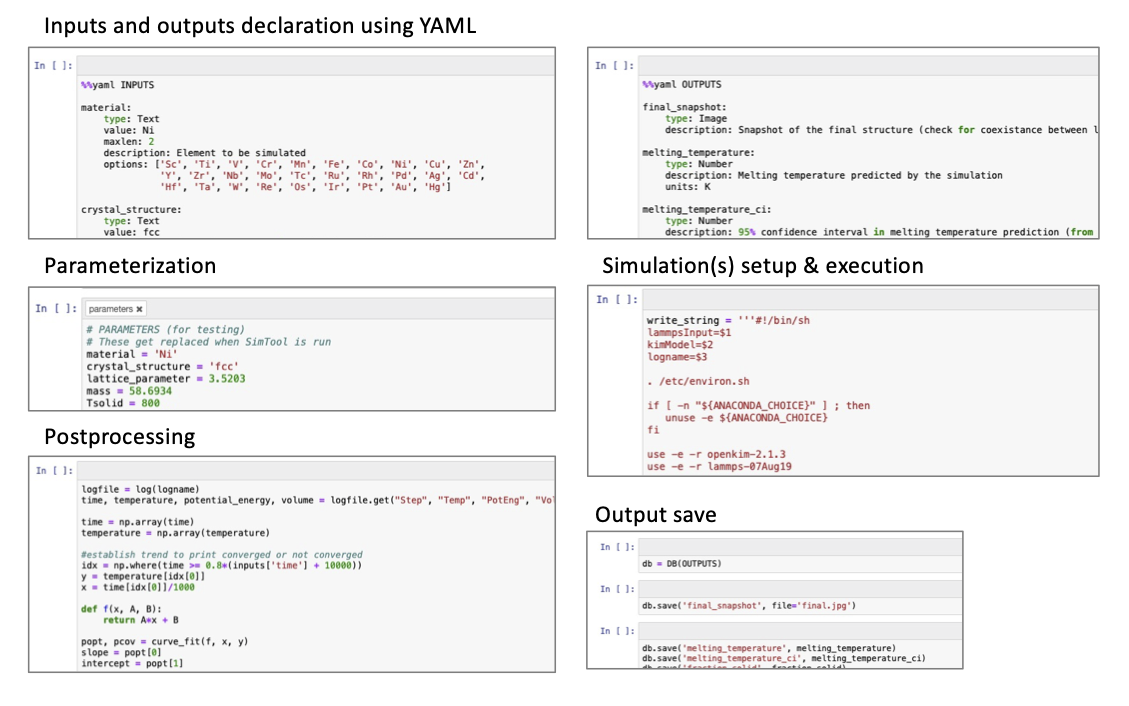}
    \caption{Elements of a \ST}
    \label{fig:SimToolEelements}
\end{figure}

\STs{} are developed and stored in Jupyter notebooks. As depicted in Fig. \ref{fig:SimToolEelements}, the main components of a \ST{} 
are: i) declaration of input and output variables using YAML \cite{papermill2017}, ii) notebook parameterization cells that use  
PaperMill \cite{papermill2017}, iii) the computational workflow connecting inputs to outputs, including all pre and 
post-processing and computations (this step can involve accessing external data resources and launching parallel 
simulation to external high-performance computing systems), and iv) population of all the output fields. 
Each element of a \ST{} is described in detail in the following paragraphs and subsection \ref{sect:simtoollib} describes the 
{\it \ST{}} library through which users interact with \STs.

{\bf Description}. The \ST{} notebook should contain a cell tagged as \texttt{DESCRIPTION}.  The plain text content of the cell 
should provide an overview of the \ST{} requirements (inputs) and services (outputs) provided, this information 
is reported when using the \ST{} library to query for available \STs{}.

{\bf Inputs}. One of the fundamental aspects of a \ST{} is that all independent input variables (those that users will 
be allowed to control) need to be declared and enumerated as a fixed list. Importantly,  developers can specify 
acceptable ranges for numerical variables. All inputs and their values are checked before execution and only
simulations with all valid inputs are accepted. \ST{} developers should decide which parameters 
will be adjustable by users and which ones will be hard-coded. The hard-coded parameters and the ranges associated with the various adjustable 
inputs should be designed to result in meaningful simulations. Importantly, by selecting the list of adjustable parameters and 
their ranges, developers can focus their \STs{} on specific tasks and minimize the chance of erroneous runs due to unphysical or
otherwise inappropriate input parameters. This is an important feature of \STs{} as most research codes do not perform such checks. 
In addition, while most scientific software has broad applicability, \STs{} enable developers to design workflows for specific 
tasks, and the explicit declaration of input and outputs enables queries to \ST{} results. As will be discussed in Section
\ref{sec:Discussion}, this is important to make the workflows and their data findable and reusable.

\STs{} accept ten types of input variables: \texttt{Boolean}, \texttt{Integer}, \texttt{Number}, \texttt{Array}, \texttt{Text}, 
\texttt{Choice}, \texttt{List}, \texttt{Dictionary}, \texttt{Image}, and \texttt{Element}.  All input types have shared characteristics: 
type, description, and value. The \texttt{Integer} and \texttt{Number} types additionally accept minimum and maximum 
values and the \texttt{Number} and \texttt{Array} types can have units. Unit conversion between user-provided input data and 
simulation input data is performed automatically using the Pint \cite{pint2019} library. 
An \texttt{Image} refers to a file using one of several popular formats including PPM, PNG, JPEG, GIF, TIFF, and BMP.  
The \texttt{Element} type allows specification of several chemical element properties using only the periodic table identifier, 
this is powered by the mendeleev \cite{mendeleev2014} library.  
The \texttt{Array}, \texttt{Text}, \texttt{List}, \texttt{Dictionary}, and \texttt{Image} types may be provided as files or Python 
variables of the proper type.  All \ST{} inputs must be enumerated in a single notebook cell using YAML.
The tool {\it Introduction to SimTools} includes all possible input types and exemplifies their use \cite{desai2020intro}.

{\bf Parameterization}. The \ST{} notebook must contain a cell tagged as \texttt{parameters}.  The cell should 
contain an assignment statement for each input. 
The example given in Fig. \ref{fig:SimToolEelements} sets specific values to the input variables, this is 
useful for developers during testing. The function \texttt{getValidatedInputs} from the \ST{} library should
be used in the parameterization cell to set default values; this is exemplified in Ref. \cite{desai2020intro}. 
When using the \ST{} library to execute a simulation 
the parameter values will be replaced by those provided by the user.

{\bf Workflow}. Following the \texttt{parameters} cell, \STs{} should include the workflow required to generate the outputs 
(described below) from the inputs. This workflow can include multiple simulations, including parallel runs in HPC resources. 
Within nanoHUB the {\it submit} command \cite{mclennan2013bringing} enables users to launch simulations in various venues
outside the execution host that powers the notebook. Importantly, this workflow should contain all the pre- and post-
processing steps required to turn inputs into outputs. While these steps are often considered unimportant and
poorly described in many publications, they can significantly affect results. \cite{alzate2018uncertainties}

{\bf Outputs}. Another key aspect of a \ST{} is that all outputs of interest must be enumerated as a fixed list.  
It should be noted that there is a difference between a \ST{} output and the simulation results.  A scientific application may produce 
many more results than what is reported by a \ST{} as outputs.  Like inputs, outputs are not optional, if an output is 
declared it must be saved during the simulation or the \ST{} library will return an error.  Output types are the 
same as the ten input types described above.  All \ST{} outputs must be enumerated in a single notebook cell using 
YAML. Developers might be tempted to include important outputs in files with ad hoc formatting. This is discouraged as it 
precludes the results from being discoverable and querieable and hinders the re-use of simulations.

{\bf Files}. The \ST{} notebook may contain an optional cell tagged \texttt{FILES}.  The cell contains a list of auxiliary files
required by the \ST{} notebook.  Examples would be additional Python files containing utility methods to support the 
simulation. In some cases it might be useful to provide files containing static data.

\subsection{Interacting with Sim2Ls: Sim2L library} 
\label{sect:simtoollib}

\begin{figure}[htp]
    \centering
    \includegraphics[width=15cm]{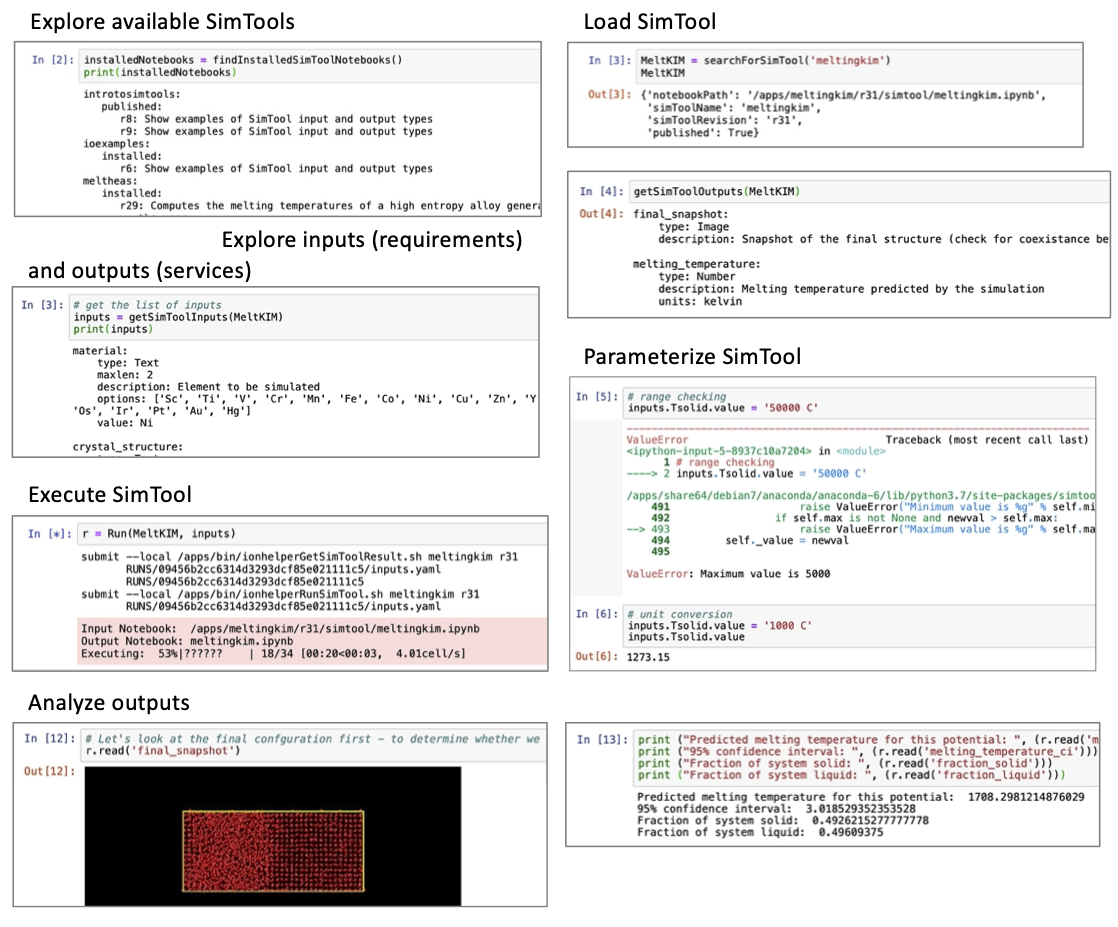}
    \caption{Discovering, parameterizing, executing, and analyzing \STs.}
    \label{fig:SimToolInvoke}
\end{figure}

Users and developers interact with \STs{} using the \ST{} library, see Fig. 
\ref{fig:SimToolInvoke}. This library 
enables users to find deployed \STs{}, their requirements (inputs), and services (outputs); it also provides a 
mechanism to executing them. 

{\bf Exploring \STs{} and setting up runs}. The \texttt{findSimTools} command enables users to discover available
\STs{} and descriptions. This command can be combined with the \texttt{getSimToolInputs} and \texttt{getSimToolOutputs} to find
\STs{} that provide the services of interest. 

The \ST{} library also facilitates simulation by providing an object used to declare all required 
inputs. This object is passed back to the \ST{} library for parameterization and execution.  Upon 
completion of the simulation a second object gives access to all declared simulation outputs.  After the successful execution 
of a \ST{}, the resulting notebook (including all inputs and outputs) is automatically stored in nanoHUB's simulation cache. 

When a new \ST{} run is requested, the \ST{} library checks the cache before execution. If a perfect match is found, the \ST{} 
library pulls the result from the cache. This not only saves compute cycles (with the consequent energy savings and reduction 
in carbon footprint) but also provides users with results nearly instantaneously. The simulation cache is particularly useful 
for computationally intensive tools and for classroom use when many users perform identical simulations.

{\bf Execution.} The papermill \cite{papermill2017} library is used to execute the code contained in the \ST{} notebook.  
The constrained nature of \STs{} means that only the \ST{} notebook, self declared additional files, and optional user provided 
input files need be provided to run a simulation.  This well-defined structure lends itself to being able to run simulations in 
a variety of venues.  By default, simulations are executed within the HUBzero tool session environment \cite{mclennan2013bringing}.  Another 
option is to build Docker or Singularity containers that mimic the HUBzero environment.  Such containers can then be 
distributed to other locations and executed.  This strategy is used to execute \STs{} utilizing MPI or other parallel 
computational methods. The use of off-site execution utilizes the submit command and requires only minimal additional 
specification including maximum wall time and number of cores to be provided. The use of containers allows our team to 
deploy simulation execution to various resources without modifying the \ST{} itself, eliminating the need for developer customization. 

The following lists the \ST{} library functions available to interact with \STs. 
\begin{itemize}
  \item \texttt{findSimTools} - find all installed and published \STs. In addition to name and revision a brief description is returned for each \ST.
  \item \texttt{searchForSimTool} - search for a particular \ST. The search may include a specific revision requirement. 
  \item \texttt{getSimToolInputs} - get definition of each input for given \ST. Definition includes name and type for each variable plus type dependent information such as units, minimum, maximum, description, default value.
  \item \texttt{getSimToolOutputs} - get definition of each output for given \ST. Definition includes name and type for each variable plus type dependent information such as units and description.
  \item \texttt{Run} - method to run specific \ST{} with provided input values.  In addition more information may be provided if the simulation is to be executed remotely.  There is also an option to control data exchange with the results cache. 
\end{itemize}

\subsection{Publishing a Sim2L, simulation caching, and results database}

Once tested by developers the process of tool publication makes them available to nanoHUB users. Every published nanoHUB tool
is assigned a digital object identifier (DOI) which is updated as new versions are released. The tool publication 
process enables developers to specify authorship, acknowledgments, provide appropriate references, and describe the tool.
Optional supporting material can be included with the tool. Once published, nanoHUB tools, including \STs{}, are indexed by
Google Scholar and Web of Science. 
Published \STs{} can be invoked by users from any Jupyter notebook running in nanoHUB which enables them to be invoked in
high-throughput or machine learning workflows called Apps, see Section \ref{sect:examples}. 

As mentioned above, every successful \ST{} run performed on behalf of users is stored in nanoHUB's simulation cache and
the \ST{} outputs indexed and stored in the results database (resultsDB). Thus, when a user requests a simulation previously
performed it is retrieved from the cache. This results in faster response time for the user and saves computational resources.
Finally, the resultsDB can be queried using an API. Thus, every \ST{} performed in nanoHUB is automatically
stored and the results are queryable. 

\section{Sim2L examples}
\label{sect:examples}
\subsection{Melting temperature calculations using molecular dynamics}

The {\it Melting point simulation using OpenKIM} \ST{} \cite{30193} in nanoHUB calculates the melting temperature of metals 
using molecular dynamics simulations. Users specify the element of interest, the model to describe atomic interactions, and additional 
simulation parameters and the \ST{} calculates the melting temperature of the material of choice using the two-phase coexistence method \cite{morris1994melting}. 
In this approach one seeks to achieve the coexistence between a liquid and a crystal phase, by definition the temperature at which this 
occurs is the melting temperature of the system. The tool creates a simulation cell, assigns initial
temperature values to two halves of the simulation box and, after a short equilibration, performs a molecular 
dynamics simulation under constant pressure and enthalpy. The choice of ensemble results in the system temperature naturally evolving 
towards the melting temperature and if coexistence is observed once the system reaches steady-state, the system temperature corresponds 
to the melting
temperature. If the entire cell ends up as a solid, the initial temperatures were too low and should be adjusted upward.
Conversely, if the entire system melts, the initial temperatures were too high. The \ST{} sets 
up, executes, and analyzes the simulation results. The simulation reports the fraction of solid and liquid phases, 
the time evolution of the instantaneous temperature and the overall system temperature with a confidence interval. 
In addition, the \ST{} analyzes the data to report whether a meaningful melting temperature can be extracted from the 
simulation. Below is a description of the key inputs and outputs, focusing on the use of the
\ST{} library to standardize this melting point calculation protocol. The \ST{} is available for online
simulation in nanoHUB \cite{30193}.

\subsubsection{Inputs}

{\bf Material}: Users input the element for which they wish to calculate the melting temperature. This input can be one of 29 metals, listed explicitly in the `options' keyword of the \ST{} input. This explicit listing allows users to quickly inspect this \ST{} input and determine the list of allowed elements. The complete list of elements can be found in Ref.\cite{30193}. 

{\bf Mass}: The \ST{} requires the atomic mass of the material as an input, this is of type `Element'. This allows users to 
either specify a numeric mass value or simply specify the symbol of the element, which the \ST{} library 
uses to automatically obtain the mass.

{\bf Crystal Structure and Lattice Parameter}: The crystal structure can be specified to be face centered cubic (FCC), body centered cubic (BCC) or hexagonal close packed (HCP). The `options' feature for this input  prevents users from selecting 
any other crystal. The \ST{} expects the lattice parameter to be a number between 2 and 10 Å. However, by leveraging the 
Pint unit conversion tool \cite{pint2019}, the \ST{} library allows users to specify the 
lattice parameter in any units. The \ST{} library automatically handles unit conversion and checks whether the converted value belongs to the range of the \ST{} input. Thus, a user input of `0.5 nm' is automatically converted to 5 Å, but a user input of `5 nm' will result in 
an error as the input is internally converted to 50 Å, beyond the range allowed by the \ST{}.

{\bf Solid and Liquid Temperatures}: The \ST{} inputs also include the initial temperatures to assign to the solid and the
liquid regions of the simulation. Users can enter temperatures in any units as the \ST{} library automatically converts 
them to Kelvin.

{\bf Run Time}: Users can also specify the time for which the coexistence calculation is carried out. The run time 
determines whether the simulation is converged or not. Short run times can result in non-steady state conditions
and unreliable calculations. 
The run-time is also internally converted to femtoseconds, with a default of 50000 fs or 50 ps.

{\bf Interatomic Model}: Every molecular dynamics simulation requires an interatomic model to describe the interactions 
between atoms. The {\it meltingkim} \ST{} obtains the user-specified interatomic model from the OpenKIM repository \cite{tadmor2011potential}. 

\subsubsection{Workflow and outputs}

The \ST{} takes in all the user inputs, creates an input file for the parallel molecular dynamics code LAMMPS 
\cite{plimpton1995fast}, executes the simulations, and post-processes the results to determine the melting temperature
if the simulation was successful. We describe the workflow is some detail to exemplify the various steps and
decisions required, even for a relatively simple and standard calculations. The \ST{} documents all these steps
facilitating reproducibility and accelerating progress as researchers can re-use parts or the entirity of the workflow.

The \ST{} first creates a system with the requested crystal structure and lattice parameter and initializes the solid and 
liquid with atomic velocities matching the specified temperatures. The user specified OpenKIM interatomic model is then downloaded 
from OpenKIM using their API. The KIM model name is included in the LAMMPS input file such that OpenKIM can interface 
with LAMMPS and modify any LAMMPS setting (units, atom style etc.) to run the simulation. 

Once the system is initialized, the simulation cell parameters and atomic positions are relaxed via energy
minimization. The system is then equilibrated under atmospheric pressure for 10 ps, using two independent thermostats
applied to the solid and liquid regions to keep the regions at the user-specified temperatures. 
Following the thermalization, the system is evolved via molecular dynamics under constant pressure and enthalpy (no thermostats), 
for the run time specified by the user. This phase of the simulation results either in the coexistence of solid and liquid 
phases (success) or a single phase; the latter indicates that initial temperatures need to be modified and a new
simulation must be performed.


The raw output from LAMMPS is also post-processed by the \ST{} to provide users with information about the simulation. 
Not just the systems temperature but also if both solid and liquid are coexisting in equilibrium. The final atomic configuration from the simulation is analyzed to establish whether solid-liquid coexistence
exists at the end of the simulation or the system evolved into a single phase. 
This is done by analyzing the local environment of each atom using the polyhedral template matching algorithm
\cite{larsen2016robust} as implemented in OVITO \cite{stukowski2009visualization}. Each atom is  classified into one of many crystal
structures based on its neighborhood, with any atom having an unknown neighborhood identified as liquid.

Based on this analysis a boolean output variable, `coexistence', is determined. If 35\% to 65\% of the atoms are identified to
belong to the initial crystal structure and if 35\% to 65\% of the atoms are identified as liquid, the system is deemed to have achieved 
coexistence and the output variable is set to TRUE. The \ST{} also outputs a snapshot of the final atomic configuration, 
for the users to visually inspect coexistence, see bottom panel in Fig. \ref{fig:SimToolInvoke}. 

The second test to establish a successful melting temperature the \ST{} performs is to check if the system is in equilibrium. To do this, it computes 
the the slope of the instantaneous temperature vs. time over the 20 ps of the simulation. If the absolute value of the slope is less or
equal to 10 K/ps equilibrium is declared and a second boolean variable, `steady state', is set to TRUE. Lastly, the temperature obtained from the last 20 ps of the simulation is reported as an output and fluctuations of the instantaneous 
temperature are used to determine the 95\% confidence interval on the melting temperature.

The \ST{} then saves the melting temperature, the confidence interval, the `coexistence' and `steady state' flags, and the fraction of atoms belonging to each crystal structure. This is performed using the save() command from the \ST{} library that allows these results to be stored in the Results Database, for easy access later.

\subsubsection{Invoking the Sim2L and example results}

The tool \cite{30193} also contains a Jupyter notebook to invoke the \ST and
demonstrate its use. This driver notebook exemplifies the use of \texttt{getSimToolInputs()} and \texttt{getSimToolOutputs()} 
functions to understand the \ST{} inputs and outputs, following which the user specifies some or all of the inputs. 
For unspecified inputs, the \ST{} uses default values, which are also displayed when the \texttt{getSimToolInputs()} function 
is called. The Run() function invokes the  \ST{} by passing it all user inputs, which the \ST{} then uses to launch the the 
LAMMPS simulation and save the outputs to the Results Database. The \texttt{getResultSummary()} function can then be used to 
get a dataframe with the results from the simulation. 

The workflow notebook additionally showcases an example of using the \ST{} to calculate the melting temperature of a 
list of elements in an automated manner. We define functions to query repositories such as Pymatgen \cite{ong2013python} for 
elemental properties to be passed as inputs. We also query OpenKIM to find interatomic models appropriate for the 
element. This example demonstrates how using the \ST{} as a fundamental compute unit can help users develop complex 
workflows and script multiple runs, while utilizing \ST{} library capabilities such as unit conversion, input validation, 
and result caching.

\begin{figure}[htp]
    \centering
    \includegraphics[width=15cm]{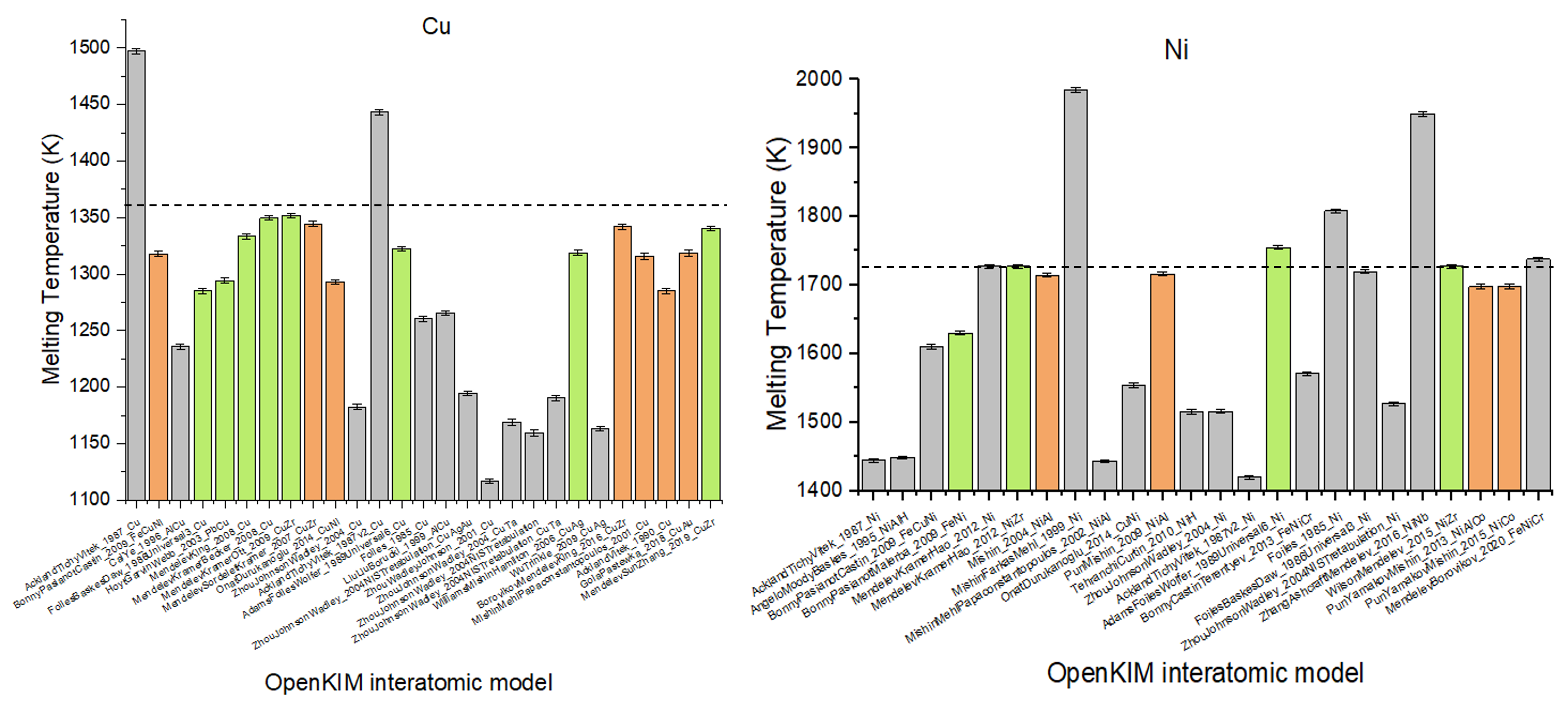}
    \caption{Melting temperature for copper and nickel, using all OpenKIM interatomic models (bars) compared to experimental results (dashed lines). Green bars represent calculations which achieved coexistence and steady state, orange bars are calculations which achieved coexistence but not steady state, indicating that longer run times can successfully determine the melting temperature. Gray bars are calculations that did not result in coexistence}
    \label{fig:melttemps}
\end{figure}

As an illustration of this capability, Figure \ref{fig:melttemps} shows the predicted melting temperatures for copper and nickel using all the interatomic models available for that metal on the OpenKIM repository. Each bar shows the melting temperature predicted for a particular model, with the error bar indicating the uncertainty in the calculation.

\subsection{P-N junction diode properties using semiconductor device simulations}

The {\it P-N junction} \ST{} \cite{mejia2021pnjunction} uses the device simulator PADRE 
(Pisces And Device REplacement software)\cite{pinto_1993} to explore basic 
concepts of P-N junctions. A P-N junction consists of a P-doped and an N-doped semiconductor arranged in series and has the electrical characteristics
of a diode. Despite their simplicity, their operation involves several fundamental concepts of solid state physics and these
devices provide useful pedagogical examples and are at the heart of many electronic devices. The \ST{} models these devices solving a coupled 
set of partial differential equations describing its electrostatics (Poisson equation), drift-diffusion (carrier continuity equation), 
and energy balance (carrier temperature equation). Users of the \ST{} can change doping concentrations for each section of the device, modify materials, 
the operating temperature, and tune additional properties and explore the resulting I-V characteristics, electronic 
band structures and hole/electron recombination. The \ST{} verifies the input parameters, creates the required input 
files required for PADRE, and stores energy bands, carrier densities, net charge distribution, voltage-current (IV) 
characteristic, and other properties as output variables. 

\subsubsection{Inputs}

{\bf Dimensions}: The \ST{} inputs include dimensions of each section in the device, P/N doped and intrinsic regions. PADRE expects 
these values in microns, however, \ST{} users can use any unit that represents distance, the \ST{} library process the values and transform the values accordingly.

{\bf Mesh refinement}: The \ST{} also expects values for the meshing required by the regions mentioned before, all values are expected to be positive and dimensionless.

{\bf Doping concentration}: Doping levels are required for the P/N type regions, and values are expected on \(cm^{-3}\) units, these values can be expressed on any scientific notation supported by YAML.

{\bf Material}: The material properties used for the simulation depend on the parameter passed to the \ST{}, the material input is a 
string, and supports selected semiconductors (`Si', `Ge', `GaAs', and `InP').

{\bf Additional Inputs}: The \ST{} also expects inputs for temperature, carriers lifetime, applied voltage, intrinsic region impurity, and environmental options. All the parameters, units, ranges and restrictions are defined on YAML on the cell tagged as "INPUTS".

\subsubsection{Workflow and Outputs}

The \ST{} translates inputs as an ASCII text file required by PADRE to run the charge transport analysis.  PADRE  can calculate  DC,  AC  small signal,  and transient solutions. The input file generated defines the structure, material models, numerical methods, and solutions. Meshes for each region of the device are defined based on the length and doping level of each region. The transport model includes Shockley-Read-Hall generation/recombination process, concentration-dependent mobility model,  field-dependent mobility, and impact ionization process.  The \ST{} first calculates the solution for the equilibrium state, and then the solutions are calculated for bias applied to the anode. The bias is increased on the specified range, and the step size defined by the inputs. PADRE's outputs are saved as text files, files are post-processed and the results saved as the Sim2L outputs. \ST{} outputs provide users with the characteristics and quantities representative of the device. The most relevant outputs of the \ST{} are described next.

{\bf Energy Bands}: The \ST{} calculates electron and hole energies, conduction band (Ec), valence band (Ev), intrinsic Fermi energy (Ei), 
and Fermi levels along the dimension of the device. Together these outputs represent the band diagram that describes the operation of the 
device under the desired conditions. The \ST{} not only calculates energies at equilibrium but also under different bias potentials. 
This can be used to visualize evolution of the band diagram as voltage is increased. 

{\bf Device Characteristics and related outputs}: The \ST{} calculates and outputs current-voltage characteristics as well as capacitance.
In addition, doping densities, electric fields, charge densities, potentials and recombination 
rates as function of position are tool outputs.

\subsubsection{Running the Sim2L via an App}

\STs{} can be invoked from Python scripts, including Jupyter notebooks, or from graphical user interfaces. 
The P-N junction tool includes an easy-to-use GUI implemented in a Jupyter notebook \cite{mejia2021pnjunction}. 
This App enables users to set inputs and visualize the device band structure, recombination rates,
as well other \ST{} outputs. The workflow within the App calculates the electric field and potentials 
using the depletion approximation. This approximation  assumes that the depletion region around the junction 
has well-defined edges and transitions between regions are abrupt. 
The workflow only includes approximations for junctions without intrinsic regions, in equilibrium, and only use the ideal Silicon intrinsic doping. The approximation is displayed together with the simulation results for educational purposes.

\begin{figure}
\includegraphics[width=.9\linewidth]{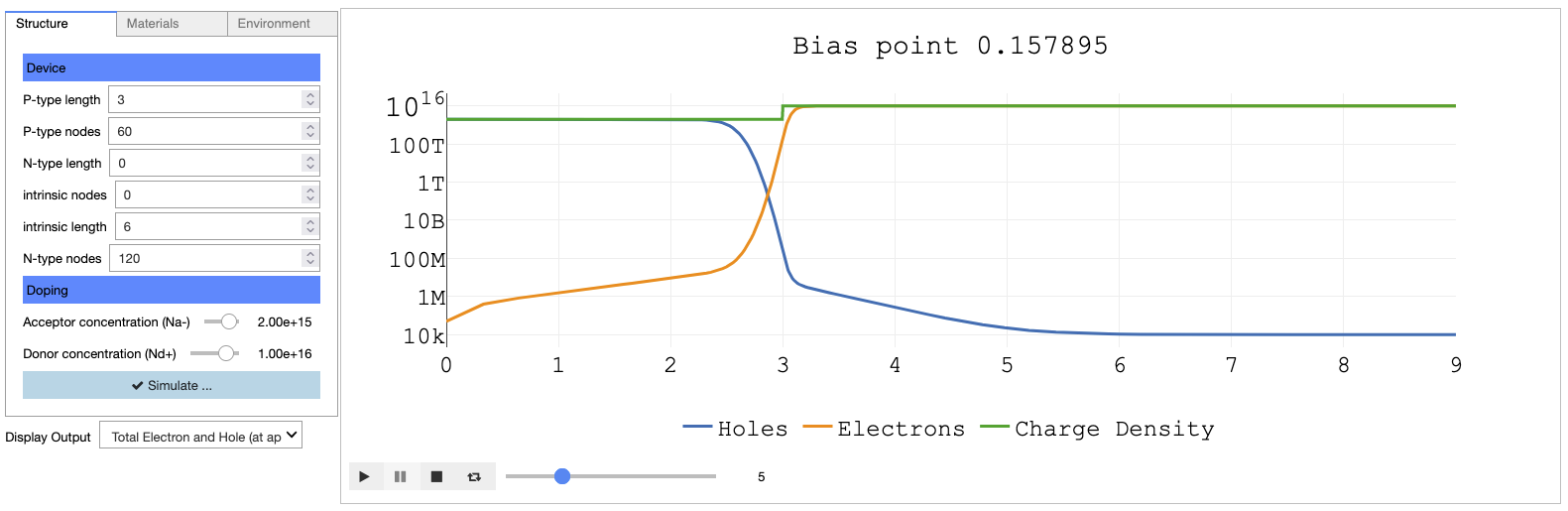}
    \captionof{figure}{P-N junction App invoked a \ST{} that performs the device simulations. The App enables users to
    easily setup the simulation and visualize results. The example shows the charge density of the diode for an applied voltage of 0.157895 eV, the slider allows user to visualize different applied voltages }
    \label{fig:simtool}
\end{figure}

\section{Discussion and outlook}
\label{sec:Discussion}

This section discusses important aspects of the simulation ecosystem for developers to consider when designing
\STs{}. While nanoHUB makes \STs{} and their data automatically accessible (via DOIs, standard licenses and APIs), these 
additional considerations are important to facilitate findability, interoperability, and reuse and the \STs{} themselves and the
data they produce.

{\bf Descriptions and metadata.} \ST{} abstracts are required as part of the publication process and the \ST{} itself has
a [description] field that can be queried when searching for \STs{}. 
Detailed descriptions help users find the appropriate tools. In addition, concise and accurate descriptions of 
inputs (requirements) and outputs (services) help with findability. 

{\bf Narrow focus vs. general Sim2Ls.} We believe narrowly defined \STs{}, i.e. workflows designed to accomplish 
specific tasks, contribute to the usability of the tool and the findability and reuse of the results produced. 
The success of large repositories of ab initio materials data is due, at least in part,
to the specific nature of the quantities included.\cite{jain2016research}

Many physics-based simulation codes have a very broad applicability, and \STs{} can be used to establish workflows 
for specific tasks. For example, molecular dynamics simulations can be used to explore mechanical properties, 
chemical reactions, shock physics, thermal transport, in materials ranging from metals to bio-inspired composites. 
\STs{} can be used by researchers in all those fields to document and share specific workflows targeting different 
properties.

{\bf Input and outputs.} The choice of inputs and outputs and their descriptions is critical to make \STs{} and their
data FAIR. While files are allowed as input and outputs, their use should be very limited since they can defeat the 
purpose of queriable inputs, outputs, and results. For example, a \ST{} could take the input file of a 
physics-simulator as the only input and produce a single output that contains a tar file of all results. 
This is strongly discouraged. Inclusion of results files from the simulator as a \ST{} output 
{\it in addition} to outputs that focus on the quantities of interest may be useful to enable users to perform
a detailed exploration of their runs and even identify problems with certain simulations. 
Another acceptable use of output files are well defined file types like PDB for molecular structures or CIF files 
for crystals.

{\bf The results database (ResultsDB).} All cache simulations in nanoHUB are indexed and stored in the ResultsDB 
and can be explored via an easy-to-use API\cite{DBExplorerAPI}; these elements will be described in a subsequent 
publication. The ability to query and re-use community-generated \ST{} results highlights the importance of 
carefully defining inputs and outputs quantities and types and designing complete end-to-end workflows that
generate all relevant quantities of interest. 

{\bf Web services}. \STs{} can be launched from within the nanoHUB simulation environment, either a terminal or 
a Jupyter notebook. In addition, \STs{} can be queries and launched from nanoHUB web services as will be described
subsequently.

\section{Conclusions}

In summary, \STs{} are a key component of the nanoHUB ecosystem to deliver simulations and their data. 
Queryable descriptions, requirements, and services (including metadata) and the use of standard technologies
make both the workflows and data FAIR. 
The declaration of inputs and outputs, including metadata, together with the simulation cache and ResultsDB means 
that all data generated can be explored, analyzed, and repurposed. \STs{} are available in the open platform 
nanoHUB both for developers and users. nanoHUB provides a complete scientific software development environment 
and compute power free of charge and online to lower the barrier of access to advanced simulations and to level 
the playing field in computational science.

\section{Data availability}

All data generated from the use of \STs{} is automatically cached by nanoHUB and indexed in the ResultsDB
that can be queried by all nanoHUB users \cite{DBExplorerAPI}. nanoHUB account are free and can be opened at: 
\url{https://nanohub.org/register/}

\section{Code availability}

The \ST{} library is available for online simulation in the open platform nanoHUB \url{https://nanohub.org},
and for download at \url{https://github.com/hubzero/simtool}. 
Documentation is available at \url{https://simtool.readthedocs.io/en/stable/}.

\section{Acknowledgements}
This work was partially supported by the Network for Computational Nanotechnology, a project of the US National 
Science Foundation, EEC-1227110. Stimulating discussions with Michael Zentner and Gerhard Klimeck are 
gratefully acknowledged.

\section{Author Contributions}

\section{Competing Interests}

The author(s) declare no competing interests.

\bibliography{Library.bib}

\end{document}